\documentclass[nofootinbib]{revtex4}
\usepackage{amsmath}
\usepackage{dcolumn}
\usepackage{bm}
\usepackage{xcolor}
\usepackage{natbib}
\usepackage{url}
\usepackage{hyperref}
\usepackage{graphicx}
\usepackage{float}

\begin{document}

\title{Testing $\Lambda$CDM cosmology in a binned universe: anomalies in the deceleration parameter}

\author{Erick Pastén}
\email{erick.contreras[at]postgrado.uv.cl}

\author{V\'ictor H. C\'ardenas}
\email{victor.cardenas[at]uv.cl}

\affiliation{Instituto de F\'{\i}sica y Astronom\'ia, Universidad de
Valpara\'iso, Gran Breta\~na 1111, Valpara\'iso, Chile}


\begin{abstract}
We study the reconstructed deceleration parameter splitting the data in different redshift bins, fitting both a cosmographic luminosity distance and also assuming a flat $\Lambda$CDM model, using the Pantheon+ sample of type Ia supernova data (SNIA). We observe tensions $\sim 2\sigma-3\sigma$ for different redshift and distance indicators if the full sample is used. However, those tensions disappear when the SNIA at $z<0.008$ are removed. If the data is splitted in 2 hemispheres according to our movement w.r.t CMB, a strange $3.8 \sigma$ tension appears in one of the samples between particular redshift bins. Finally, considering posterior distribution as Gaussian, general linear model prefers a positive slope for $q_0$ across redshift bins opposed to a zero slope expected in a $\Lambda$CDM universe. We discuss possible explanations for our results and the influence of lowest redshift SNIA data in cosmological analysis.
\end{abstract}


\maketitle


\section{Introduction}\label{section:1}

$\Lambda$CDM is the current standard model in physical cosmology, assuming a unique set of parameters that governs the evolution of the universe at the present epoch. According to studies using the last \textit{Pantheon}+ SNIA compilation \cite{Scolnic:2021amr}, \cite{Brout:2022vxf}, a non zero cosmological constant $\Lambda$ is still preferred, leading to the scenario of an accelerated universe with a negative deceleration parameter $q_0$ today. However, deeper studies are necessary to extract information of the actual nature of dark energy (DE) from the data.

Cosmic acceleration of the universe is one of the strongest and most important astrophysical discoveries in the last years \cite{Perlmutter99}, \cite{Riess98}. The use of the FLRW metric to describe the evolution of an isotropic and homogeneous universe has been extremely successful to match observations such as cosmic microwave background (CMB), SNIA and Baryon acoustic oscillations (BAO), indicating a negative value for the ``deceleration" parameter $q_0$ explained by adding a cosmological constant $\Lambda$ in the Einstein equations \cite{Betoule14a}, \cite{Auborg15}, \cite{Baxter16} \cite{Alam17}, \cite{Efstathiou18}, \cite{Scolnic18},  \cite{Aghanim20}. This constant is ultimately related with an unknown form of energy that has been called dark energy (DE) and represent one the most important challenges for the astrophysics and physics in the XXI century (see \cite{Peebles03}, \cite{Durrer11} for reviews). Although DE is accepted as the standard explanation for the observations in cosmology, many theoretical \cite{Velten14},  \cite{Weinberg00} and observational \cite{DiValentino_2021}, \cite{Riess21a}, \cite{Perivolaropoulos:2021jda} tensions remains, and the questioning on the veracity of $\Lambda$CDM has becoming strong in the recent years.
According to ideas on inhomogeneous \cite{Celerier06}, \cite{Alexander09}, \cite{Cosmai19}, \cite{Pasten_Fractal},\cite{Clarkson_2011} and tilted cosmologies \cite{Tsagas11}\cite{Tsagas15}, the late structure of the universe could influence strongly the interpretation of cosmological data as we are not typical observers as Copernican Principle state, which is related to being located in a under/over density and/or in the middle of a bulk flow that is not in rest w.r.t the Hubble Flow. 

 Observational evidence points to the fact that locally the distribution of matter alters the measurements, causing the homogeneity scale to be even above 300 Mpc. In the context of the LCDM model, this scale indicates that under redshift 0.1 we should observe such effects. It was using type Ia supernovae that the author of \cite{Zehavi:1998gz} suggested the existence of a Hubble Bubble, the idea of a local void. 
Therefore it is reasonable not to use data at very low redshift, as in \cite{Kenworthy19} for example. However, Pantheon+ includes data at redshift as low as $z=0.0001$, so we expect to get unusual behavior. 

It is important then to study the cosmological data at different redshift and for different directions to check the cosmological principle assumptions. A key previous work is \cite{Asvesta:2022fts} where the authors show that a tilted Einstein-de Sitter universe can reproduce the recent acceleration of the universe, without any DE component, just by  simply taking into account linear effects of peculiar motions. In this paper we study the local universe using a cosmological differential approach, extracting the cosmological deceleration parameter $q_0$ for different redshift bins of SNIA data. We observe a qualitative and quantitative deviations from what is expected in a $\Lambda$CDM universe. Therefore, we study possible causes for those anomalies.

\section{Cosmographic analysis}

The well known cosmographic Taylor expansion \cite{Visser_2004,Visser_2005}up to the third term or the luminosity distance is given by:
\begin{equation}
    D_L(z)=\frac{z}{H_0}+\frac{(1-q_0)}{2H_0}z^2-\frac{(1-q_0-3q_0^2+p
    _0)}{6H_0}z^3,
\end{equation}
with $a(t)$ being the scale factor of the FLRW universe, $q_0$ is the \textit{deceleration} parameter $q=-\ddot{a}{a}/\dot{a}^2$ evaluated today and $p_0=j_0-\Omega_k$, where $j=\dot{\ddot{a}}/aH^3$ is the \textit{jerk} which is degenerated with the curvature parameter and cannot be determined separately. Usually it is argued that $q_0 \sim -0.5$ from supernovae data \cite{Scolnic_2018} and this is interpreted as an acceleration of the universe due to the presence of the mysterious DE. However, this interpretation changes if we consider inhomogeneous or tilted cosmologies, in which case $q_0$ is determined by the inhomogeneties rather than a cosmological constant.

Motivated by ideas on inhomogeneous and tilted cosmologies, we perform a differential study binning the data for different redshift ranges, to determine if the best fit of $q_0$ changes with redshift. In this case we use
\begin{equation}
    D_L^{i}(z)=\frac{z}{H_{0i}}+\frac{(1-q_{0i})}{2H_{0i}}z^2-\frac{(1-q_{0i}-3q_{0i}^2+p
    _{0i})}{6H_{0i}}z^3,
\end{equation}
for the $i$-th bin in the sample. The procedure for binning SNIA data is explained in the following section. To perform a comparison, we use both the cosmographic luminosity distance as well as the one for the flat $\Lambda CDM$ model. In this case the $q_{0i}$ parameter is the same for every redshift given by:
\begin{equation}\label{eq3}
    q_{0i}=\frac{1}{2}\Omega_{mi}-\Omega_{\Lambda i}=\frac{3}{2}\Omega_{mi}-1.
\end{equation}
The main difference is that the fit of $q_0$ from the cosmographic approach needs to be performed for redshifts below $z \simeq 0.3$ as the Taylor approximation crashes at higher values, while for the fit of $q_0$ using $\Lambda$CDM we can use all the data for which the quantity of SNIA in those redshift is significant.

We have used $z \sim 0.8$ as an upper redshift limit for $\Lambda$CDM cosmology, as the number of data points per redshift bin for values larger than $z =0.8$ falls rapidly to less than $40$ data points per bin, very low to perform significant statistical analysis. We hope that this redshift threshold increase in the future.

It is important to notice that the cosmographic approach is commonly described as a \textit{fully independent model relation}, but that is not exactly true. This relation comes from the assumption that we are co-moving observers with the so-called CMB rest-frame (which is also assumed to be in rest with respect to the Hubble Flow). This means that possible effects of our relative motion are reduced to redshift corrections on observer and the data due to peculiar velocities.

If the data is corrected due to peculiar velocities, corrections has to been made also over the luminosity distance \cite{Davis11}. However, as pointed out in some studies \cite{Tsagas11,Tsagas15}, strong cosmological implications could be hidden in this relative motion more than just data corrections. If $z_{\odot}$ is the peculiar movement of our solar system moving w.r.t. the CMB and $z_{sn}$ is the peculiar movement of each supernovae, then the luminosity distance should be corrected to: 
\begin{equation}
    \bar{D}_L(z,z_{\odot},z_{sn})=(1+z_{sn})^2 (1+z_{\odot})D_L(z),
\end{equation}
where $\bar{D}_L$ is the corrected luminosity distance and $z$ is the cosmological redshift. This redshift corrections should not modify significantly the cosmological parameters estimation \cite{Davis_2019}. However, as we are doing a redshift binned analysis, those corrections could be more important at low redshift bins.

\subsection{SNIA analysis: The Pantheon + Sample} 

The Pantheon + sample \cite{Scolnic:2021amr}, \cite{Brout:2022vxf} is the latest compilation of high and low redshifts SNIA and has a lot of improvements on photometric uncertainties and systematic as well as peculiar velocity corrections based on a velocity field reconstruction, improving the quantity of data at low redshift. To extract the information on $q_0$ from the data, we use the full Tripp Formulae for distance modulus:
\begin{equation}
    \mu_{obs}=m_b^*-M=m_b+\alpha x-\beta c+\Delta_M-M
\end{equation}
here, $m_b$ corresponds to the peak magnitude at time of B-band maximum, $M$ is the absolute B-band magnitude of a fiducial SNIA, $x$ and $c$ are light curve shape and color parameters corrections, $\alpha$ is a coefficient of the relation between luminosity and stretch, $\beta$ is a coefficient of the relation between luminosity and colour, and $\Delta_M$ is a corrections based on the mass of the host galaxy. 
This relation is used to compare with the theoretical expectation of a particular model using standard statistical analysis where:
\begin{equation}
    \mu=5\log_{10}{\frac{D_L}{10Mpc}}+25.
\end{equation}
In a $\Lambda$CDM universe we expect to obtain a uniform distribution values for $q_0$ as any part of a homogeneous-isotropic universe should lead to the same `` today" deceleration parameter. This expectation increases as the corrected magnitudes in Pantheon + has been performed even allowing for peculiar velocities corrections and redshift dependent nuisance parameters. As we are interested just in the $q_0$ parameter, we marginalize over the degenerated parameter involving $H_0$ and $M$ according to \cite{Conley_2010}.

\subsection{Binning}

We follow two different approaches to study the $q_0$ distribution as a function of redshift, searching for cumulative and local effects in the Pantheon+ sample. 
\begin{figure}[h!]
    \centering
    \includegraphics[width=10cm]{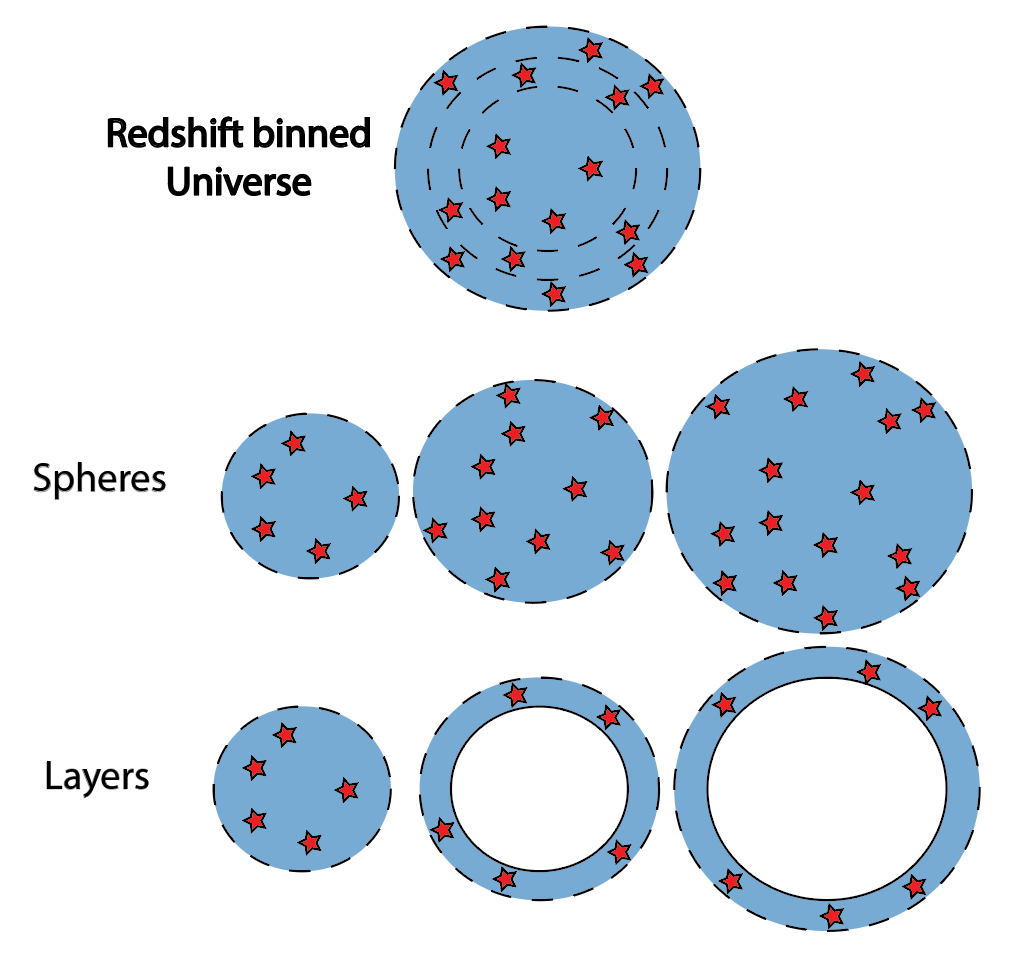}
    \caption{The 2 approaches used to describe $q0$ distribution using redshift bins.}
    \label{scheme}
\end{figure}

\subsubsection{Redshift spheres}

We begin with a sphere centered at $z=0$ of radius $z_{out}=0.01$ and we fit $q_0$ for all supernovae inside this sphere. Then, we increase the radius of the sphere in steps $\Delta z$ repeating the analysis for supernovae inside the new sphere until we reach the radius $z_{out}=0.3$. This method allows us to describe the behaviour of the local deceleration parameter and to search where DE existence becomes noticeable. Also, this procedure has the advantage that the number of data always increases from a initial value which is the number of supernovae with $z \leq 0.01$. \textbf{A similar approach was used in \cite{Colgain_2019}}. However, we are not able to describe how $q_0$ behaves locally for each redshift as this study search for cumulative effects.

It is well established the presence of a Hubble Bubble in low $z$ (see e.g. \cite{Kenworthy19}) leading to the usual removal of low redshift supernovae with $z\leq0.023$ in cosmological analysis as we can expect any unusual cosmological inference at those redshift.
 
 Because SNIA with low redshift induce a bias in the Hubble diagram we also perform our study avoiding the use of data with $z<0.023$.  Overall, it is difficult to look for effects of the local structure in cosmology if we do not consider the redshifts in which those effects are more powerful. Therefore, we first consider the analysis using all supernovae and then remove the low redshift ones to look for variations.

\subsubsection{Redshift layers}

The second approach is to perform the analysis in layers of redshift.
{\bf For example, the studies performed in \cite{Kazantzidis:2020tko,Dainotti_2021,Dainotti_22}, use this type of layer selection}. 
As the previous method, we begin with a sphere this time of radius $z_{out}=0.1$. Then we increase the radius in step of $\Delta z$ but this time we also increase the inner radius in the same step. It means that the second fit is performed in a layer of redshift $z \in [\Delta z,0.1+ \Delta z]$, the second for $z \in [2\Delta z,0.1+2\Delta z]$ and so on. This method is very useful to search for local difference in $q_0$ at different redshift. 

\section{Results}

\begin{figure}[H]
    \centering
    \includegraphics[width=14cm]{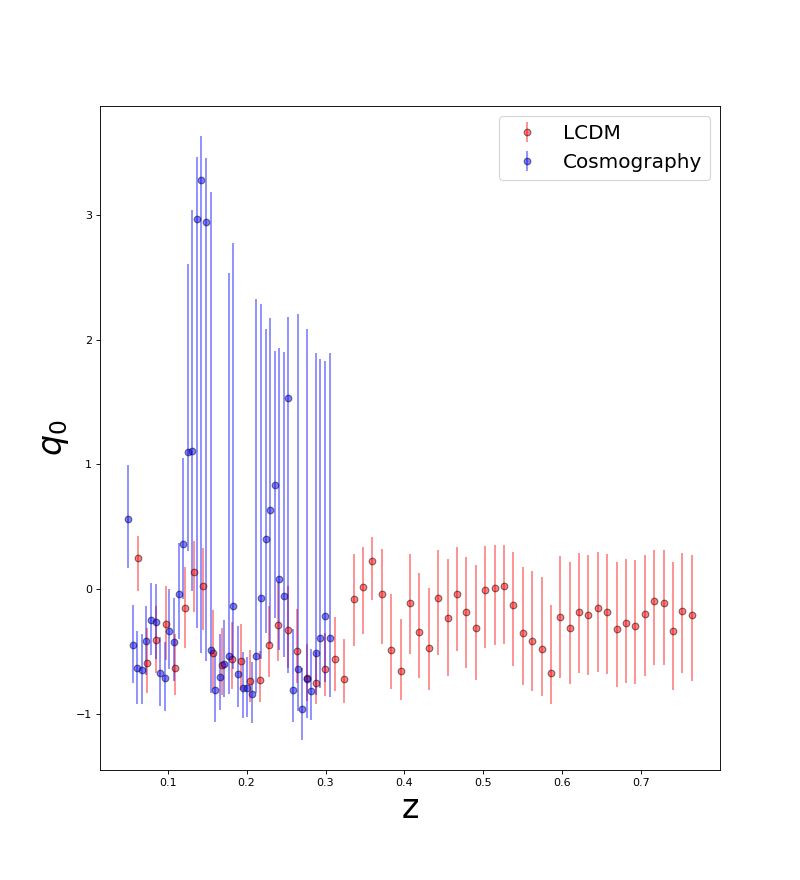}
    \caption{$q_0$ parameter using the layers approach. $z$ here is the mean redshift of the bin.}
    \label{q0varyng}
\end{figure}

\subsection{Initial Qualitative test}

\begin{figure}[H]
    \centering
    \includegraphics[width=14cm]{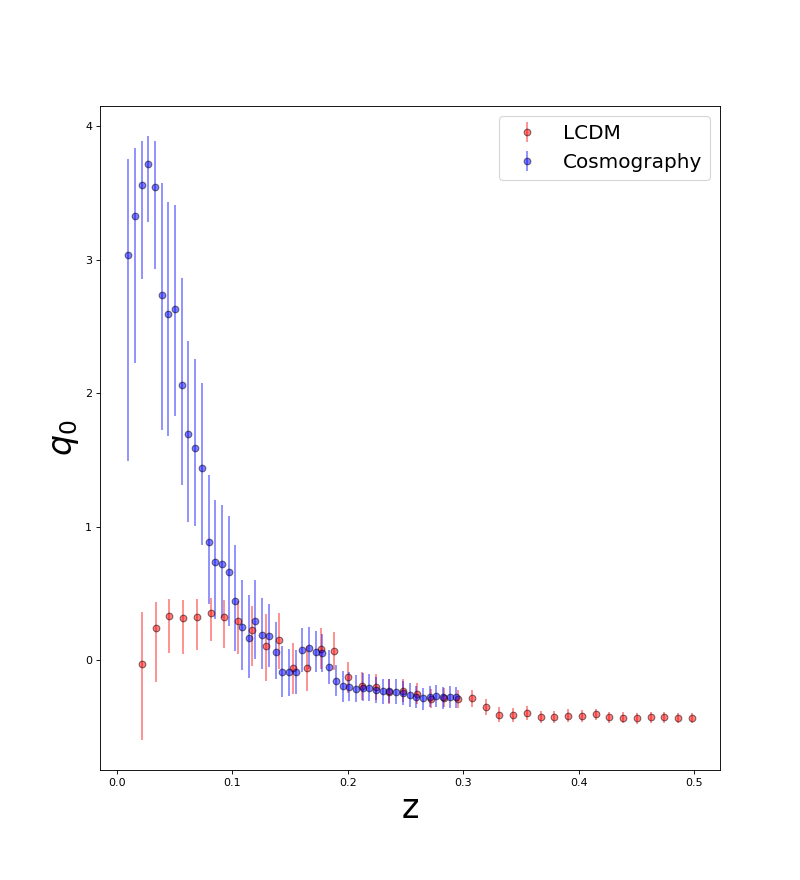}
    \caption{$q_0$ parameter using the spheres approach. $z$ here is the upper redshift of the sphere.}
    \label{q0fixed}
\end{figure}

We use the code \texttt{Emcee} \cite{2013PASP..125..306F} to test the model against the data: the Pantheon+ SNIA \cite{Scolnic:2021amr}. This is a pure python implementation of the affine invariant ensemble sampler for Markov chain Monte Carlo proposed by Goodman and Weare \cite{2010CAMCS...5...65G}. The results of fitting $q_0$ using both the cosmographic luminosity and $\Lambda$CDM are shown in Fig. (\ref{q0varyng}) and (\ref{q0fixed}) for the sphere and layer approaches respectively. To look for insights in the cosmographic approach, we use a weakly uninformative uniform prior for $q_0\sim U(-4,4)$ not restricting our analysis to any cosmology. Qualitatively, we can observe a deviation from the uniform distribution for $q_0$ expected from a $\Lambda$CDM universe. Also, it is interesting to note that without a tight prior, MCMC samples tends to prefer higher values for $q_0$ in the cosmographic Taylor approximation than the ones allowed in $\Lambda$CDM at low redshifts.

\subsection{Quantifying possible anomalies in the layer approach}

In the context of the $\Lambda$CDM model, we expect $q_0$ should take the same value independent of $z$. To quantify the possible deviation of this result, we explore by using bins in which qualitatively we see a deviation from a uniform distribution for $q_0$. We have chosen the width of each bin to show the most appreciable differences in the fit of $q_0$. The results are summarized in table \ref{tabla1}.

\begin{table}[ht]
\begin{ruledtabular}
\begin{tabular}{ccc}
 $z$ Bin & Number of SNIA & $q_0 $ \\
\hline
$0-0.15$  & 826 & $-0.087  \substack{+0.18\\-0.17}$ \\ 
$0.008-0.15$  & 746 & $-0.453 \substack{+0.16\\-0.16}$ \\ 
$0.05-0.2$ & 300 & $-0.457  \substack{0.17\\-0.16}$  \\
$0.1-0.25$ & 346  & $-0.501\substack{0.18\\-0.17}$  \\
$0.15-0.3$ & 381 & $-0.61\substack{0.16\\0.15}$ 
\end{tabular}
\caption{Values for $q_0$ fit in wide redshift bins using the cosmographic Taylor expansion of the luminosity distance.}
\label{tabla1}
\end{ruledtabular}
\end{table}

\begin{figure}[h]
    \centering
    \includegraphics[width=10cm]{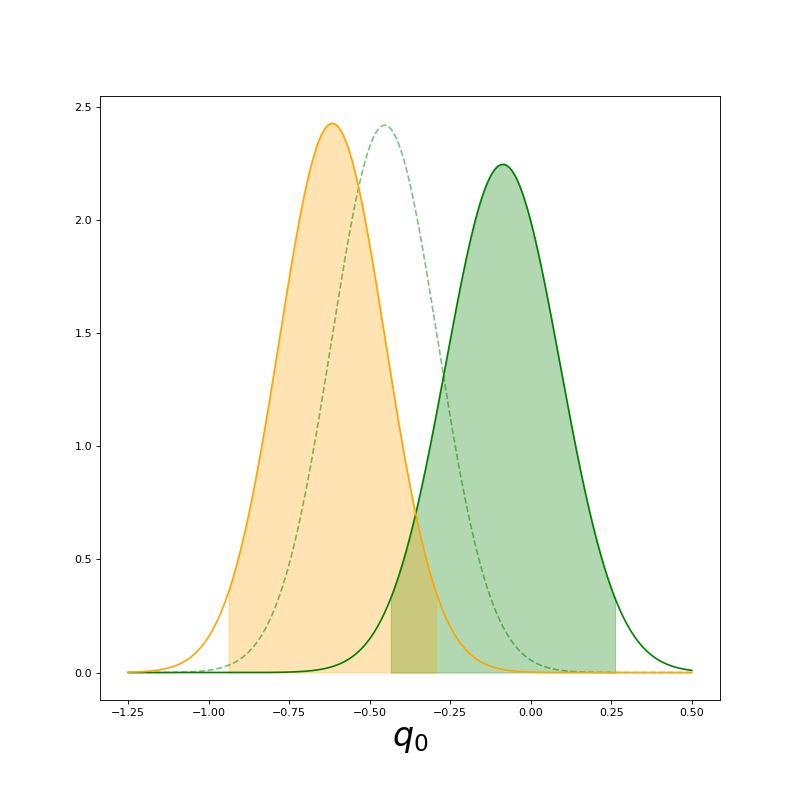}
    \caption{Gaussian distribution of $q_0$ cosmographic parameter for different redshift bins. Between $[0,0.15]$ (green) and $[0.15,0.3]$ (orange) appears a tension of $2.2 \sigma$ considering independent realizations. The tension almost disappear when supernovae below $z<0.008$ are removed, reaching a value $\sim 0.7 \sigma$ (dashed line).}
    \label{sigmas1}
\end{figure}

Assuming independent realizations, we observe that $q_0$ has a tension of $\sim 2.2 \sigma$ between the best fit values from supernovae at $z<0.15$ with those at $z\in[0.15,0.3]$. Overall as data in the two analysis are correlated, this tension could increase. However, if we remove the very low redshift SNIA data at $z<0.08$ the results change, decreasing this tension to zero with a value $\sim 0.7 \sigma$ (Figure \ref{sigmas1}).

\begin{figure}[ht]
    \centering
    \includegraphics[width=8cm]{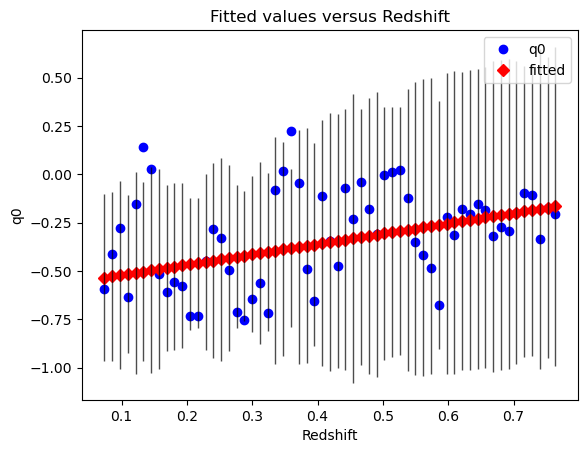}
    \includegraphics[width=8cm]{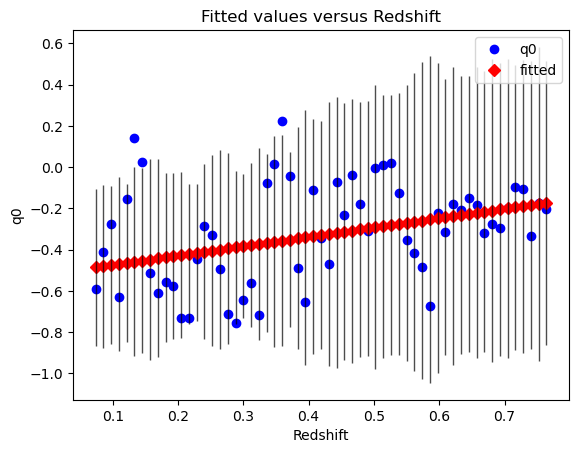}
    \caption{Best general linear model fits for $q_0(z)$ using $\Lambda$CDM luminosity distance with Gaussian error $\sigma$ estimated as the mean of the asymmetric errors from MCMC (Left) and estimated as the maximum value between them (Right).} 
    \label{Fit}
\end{figure}

To check the consistency of the derived $q_0$ values with the expectations in the $\Lambda$CDM model, we also performed a general linear model fit using the package \texttt{STATSMODELS} \cite{seabold2010statsmodels}. In this case we use:
\begin{equation}
    q_0(z)=Q_m z+Q_0,
\end{equation}
and any deviation from a zero value for $Q_m$ is an indicator of a departure from the $\Lambda$CDM model. For simplicity we approximate each fit for $q_0(z)$ in the $\Lambda$CDM as being gaussian distributed, estimating the error $\sigma$ as the average between the asymmetric errors obtained by \texttt{Emcee} and also estimating $\sigma$ as the maximum between them. We do not consider the first bin for $z \leq 0.008$ data. We also performed a fit with a zero slope $Q_m=0$ model, in order to compare with the general non zero case. Results are summarized in Table \ref{tabla2} and the best fit are plotted in figure \ref{Fit}. According to the statistics, the fit tends to prefer a non-zero slope over a constant value of $q_0$ for all redshifts. Overall, the fit is consistent with a negative mean $q_0$ parameter. 

\begin{table}[ht]
\begin{ruledtabular}
\begin{tabular}{lcc}
Statistic & Best linear fit of SNIA & Constrained fit with zero slope \\
\hline
$Q_m$ & $0.5341 \pm 0.190$ ($0.4532 \pm     0.173 $) & 0  \\ 
$Q_0$  & $-0.5736 \pm 0.071 $ ($-0.5189 \pm  0.066$)& $-0.3987      \pm 0.037(-0.3699      \pm 0.034)$ \\ 
Log-Likelihood & $-9.4073(-5.0585)$ & $-13.241(-8.4040)$  \\
Pearson Chi-Square & $35.7(26.0)$  & $40.6(29.1)$  \\
Pseudo R-Square & $0.1260(0.1101)$ & $0.0001453(0.0001453)$ \\
AIC & $22.81(14.11)$ & $28.48(18.8)$ \\
BIC & $26.96(18.27)$ & $30.55(20.89)$
\end{tabular}
\caption{Best linear fit parameters for $q_0$ using $\Lambda$CDM luminosity distance estimating the Gaussian error $\sigma$ as the average between asymmetric errors from MCMC.  In parenthesis, same values but this time estimating $\sigma$ as the maximum value between them.}
\label{tabla2}
\end{ruledtabular}
\end{table}

\subsection{Quantifying possible anomalies in the spheres approach}

\begin{figure}[ht]
    \centering
    \includegraphics[width=10cm]{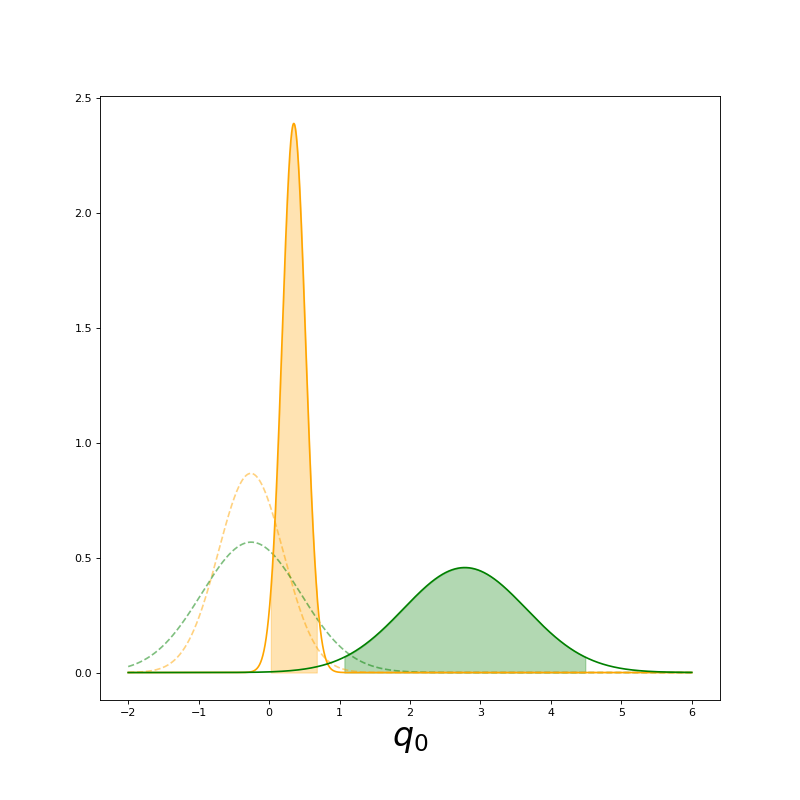}
    \caption{Gaussian distribution of $q_0$ cosmographic luminosity distance (green) and $\Lambda$CDM (orange) up to redshift $0.05$. The cosmographic distance prefer higher values for $q_0$ even when the prior is tight, giving a tension of $\sim 2.8\sigma$ between both distance indicators. This tension disappears when the low redshift $z<0.008$ data is removed (dashed line).}
     \label{sigmas2}
\end{figure}

Using the ``spheres'' method, and from the results displayed in Fig.(\ref{q0fixed}), we observe a noticeable difference between the best fit for $q_0$ from the cosmographic expansion and those from the $\Lambda$CDM model for redshift below $z =0.1$. For example, restricting our analysis to data points below $z = 0.05$, we obtain the best fit values quoted in Table (\ref{tabla3}).
Considering gaussian distributions, we obtain a tension between the cosmographic distance fit and the $\Lambda$CDM one of $\sim 2.8$ (Figure \ref{sigmas2}). Even by decreasing the parameter space for the prior $q_0 \sim U(-1,1)$, the cosmographic distance still prefer a best fit $q_0>1$. Interestingly, in both cases the best fit takes positive values of the deceleration parameter at low redshift $z<0.008$, but the value changes dramatically when they are removed, making the tension disappears.

\begin{table}[ht]
\begin{ruledtabular}
\begin{tabular}{ccc}
Distance indicator & Number of SNIA & $q_0 $  \\
\hline
Cosmographic & 648(568) & $2.78 \substack{0.98\\-0.87}(-0.25 \substack{0.73\\-0.68})$   \\ 
$\Lambda$CDM & 648(568) & $0.35  \substack{0.11\\-0.22}(-0.26  \substack{0.47\\-0.45})$  \\
\end{tabular}
\caption{Best fit values for $q_0$ in wide redshift bins using the cosmographic Taylor expansion of the luminosity distance at $z<0.05$. In parenthesis we quote the values obtained when we have removed the SNIa data with $z<0.008$}
\label{tabla3}
\end{ruledtabular}
\end{table}

\subsection{Local Structure Effects}

\begin{figure}[ht]
    \centering
    \includegraphics[width=14cm]{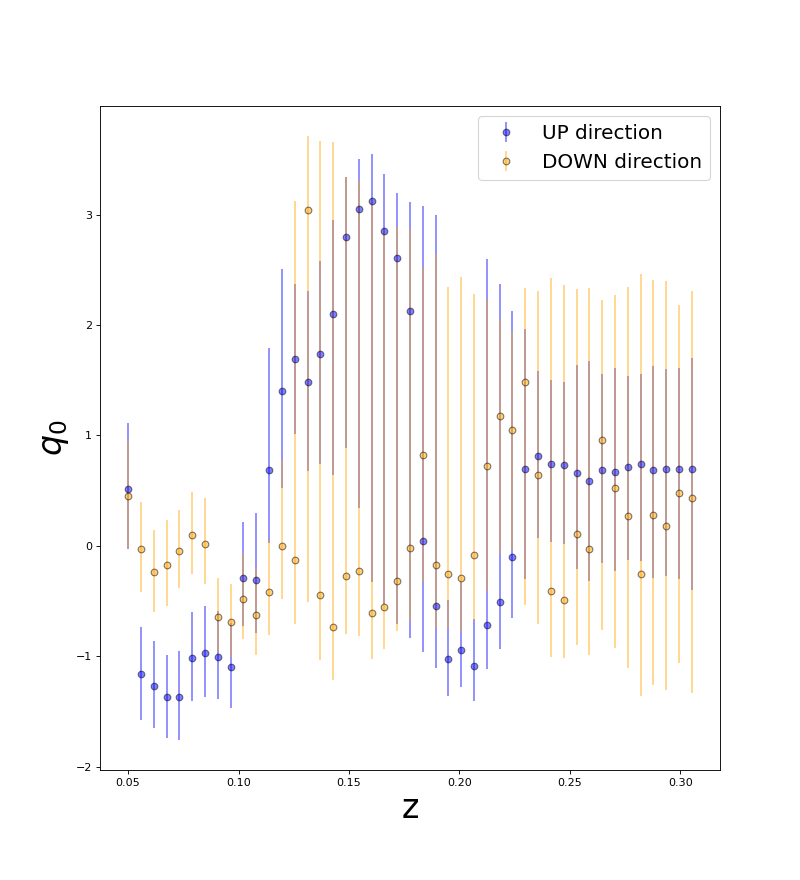}
    \caption{$q0$ dipolar parameter using the layers approach.}
    \label{q0varyngdipole}
\end{figure}

In this section we consider a possible orientation dependence of the best fit value for $q_0$ motivated by anomalies in the CMB dipole direction (see \cite{Aluri22} for a review). Similar analysis have been performed in the past with different data sets. Usually, a positive result of this orientation dependence is interpreted as a local distribution structure effect on the data. We perform a layer binned analysis with the cosmographic luminosity distance for two hemispheres, oriented to the dipole of the CMB, the one interpreted as our motion respect the CMB rest frame. To look for relations between our movement w.r.t. the CMB and the $q_0$ parameter distribution, we perform a layer binned analysis with the cosmographic luminosity distance for two hemispheres separated by our direction of movement.  We take the direction reported in \cite{Planck2020} corresponding to $(263.99,48.26)$ in galactic coordinates to split the data in two groups according to the two hemispheres of the CMB dipole. We call \textit{UP} and \textit{DOWN} respectively to the blue-shifted and red-shifted hemispheres of the dipole. Although there are others possible anomalous directions in cosmology, we selected directly the CMB dipole direction as it is the principal standard cosmological anisotropy (see \cite{Aluri_review} for a review). Also some alternative cosmologies as \textit{tilted cosmologies} predict an anisotropy in the $q_0$ parameter close to this direction (see \cite{Tsagas11, Tsagas15}).

\begin{figure}[ht]
    \centering
    \includegraphics[width=10cm]{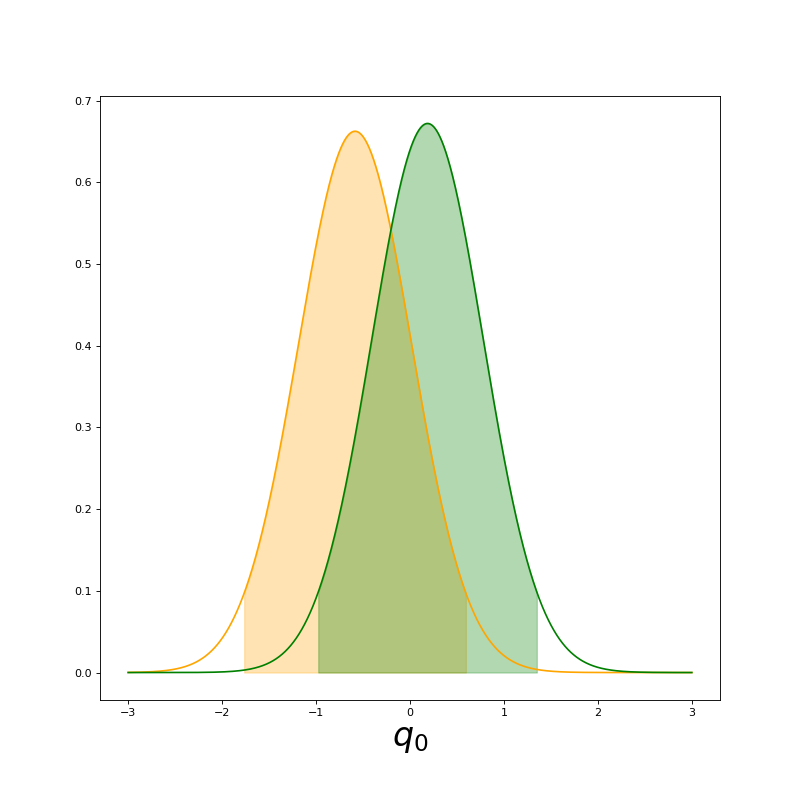}
    \caption{Gaussian distribution of $q_0$ cosmographic luminosity distance at low redshift $[0.008,0.075]$ for the 2 hemispheres. The tension between the fits doesn't reach the unity.}
    \label{sigmas3}
\end{figure}

The results are shown in Figure \ref{q0varyngdipole}. Qualitatively, we can see two possible sources of anomalies. First, a visual discrepancy of $q_0$ values between the hemispheres at low redshift is observed. To analyze this, we fit $q_0$ for the two hemispheres at $z \in [0.008,0.075]$. Results are in Table \ref{tabla4}. For Gaussian distribution of parameters, we can see that the tension between hemispheres is non-existent (Figure \ref{sigmas3}).

Notice that for $z<0.1$ the UP value for $q_0$ is lower than the DOWN one by a factor of $1$, at more than one sigma. In the range $0.11< z < 0.15$ the trend is the opposite, being the UP value higher than the DOWN one. From $z>0.2$ considering the uncertainties, both hemispheres fit  essentially the same value for $q_0$, although with larger uncertainties.

\begin{figure}[ht]
    \centering
    \includegraphics[width=10cm]{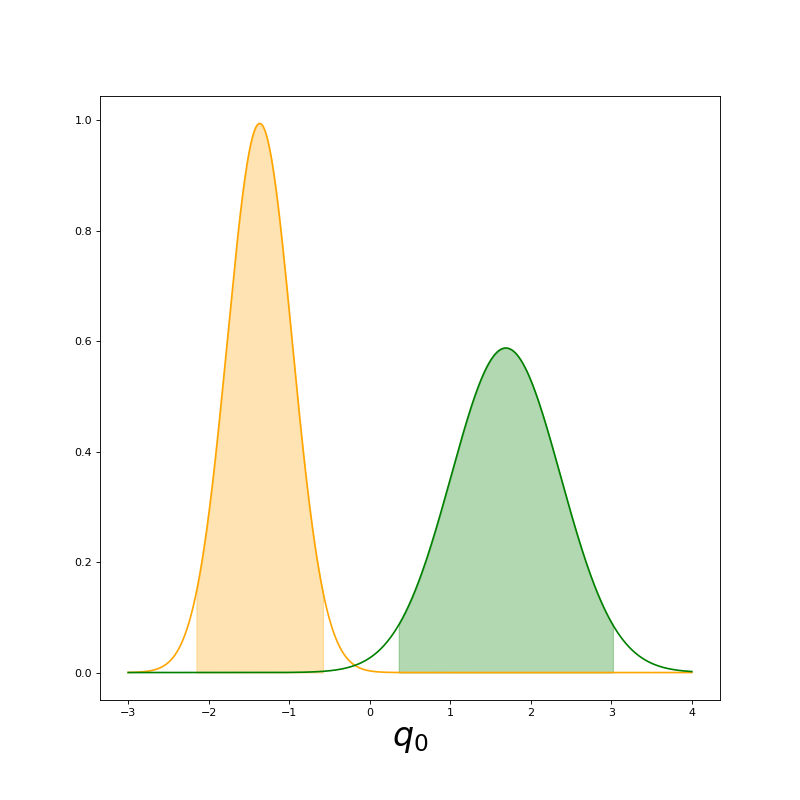}
    \caption{Gaussian distribution of $q_0$ cosmographic luminosity distance at $z \in [0.0232,0.1232]$ and $z \in [0.0754,0.1754]$ for the UP hemispheres. The tension between the posterior paremeter distributions reach 3.8 $\sigma$.}
    \label{sigmas4}
\end{figure}

Also only on the UP hemisphere, we note an increase in the $q_0$ parameter comparing the redshift ranges lower than $z\sim0.075$ and $z\sim 0.11-0.15$. For example, in particular the bins $[0.0232,0.1232]$ and $[0.0754,0.1754]$ show the best fit parameters summarized in Table \ref{tabla5}, giving a tension around $3.8 \sigma$ (Figure \ref{sigmas4}). However, at $z\sim 0.11-0.15$ there are only 23 supernovas, requiring more data to extract significant conclusions. 

\begin{table}[ht]
\begin{ruledtabular}
\begin{tabular}{ccc}
$z$ bin & Number of SNIA & $q_0 $  \\
\hline
$0.008-0.075$(UP) & 314 & $-0.59 \substack{0.62\\-0.59}$   \\ 
$0.008-0.075$(DOWN) & 314 & $0.19 \substack{0.62\\-0.57}$  \\
\end{tabular}
\caption{Values for $q_0$ fit at low redshift using the cosmographic Taylor expansion of the luminosity distance for the 2 hemispheres.}
\label{tabla4}
\end{ruledtabular}
\end{table}

\begin{table}[ht]
\begin{ruledtabular}
\begin{tabular}{ccc}
$z$ bin & Number of SNIA & $q_0 $  \\
\hline
$0.0232-0.1232$(UP) & 214 & $-1.37 \substack{ 0.41\\  -0.39}$   \\ 
$0.0754-0.1754$(UP) & 92 & $1.69 \substack{0.68\\-0.67}$  \\
\end{tabular}
\caption{Values for $q_0$ fit in particular redshift bins using the cosmographic Taylor expansion of the luminosity distance on the \textit{UP} hemisphere.}
\label{tabla5}
\end{ruledtabular}
\end{table}

\section{Discussion}

In this work we have found several unexpected behavior for the distribution of the deceleration parameter $q_0$ values  using the recent type Ia supernova sample Pantheon+.
First, analyzing the low redshift supernovae ($z<0.3$) we obtain evidence for a local decelerating universe using the cosmographic distance and considering the full sample of Pantheon+. However, this behaviour disappears once the SNIA at redshift lower than $z=0.008$ are removed. A tension of $\sim 2.2\sigma$ between the cosmographic expansion fit at different redshift bins are observed when the full sample is considered, and also a discrepancy of $\sim 2.8\sigma$ between the preferred values for $q_0$ with the cosmographic expansion function and the full $\Lambda$CDM one. Again, this behaviour disappears once the lowest redshift supernovae (those with $z<0.008$) are removed.

At this point it is interesting to comment the result of \cite{Colgain_2019} where the author find, using the Pantheon sample, the fit for $\Omega_{m0}$ varies with the maximum redshift of the data used, finding a very low value between $z \simeq 0.1$ to $z\simeq 0.15$, a result also found in \cite{Camarena_2020}, and even wiggles have been found \cite{Kazantzidis_2020}.

Also, assuming the $q_0(z)$ is a varying function 
of redshift, being described by a linear model, the statistical analysis favours a non-zero slope according to AIC and BIC criteria, showing a clear departure from the $\Lambda$CDM universe. Finally, we performed a binned study in the hemispheres defined by the CMB dipole. We did not found evidence for tensions between the hemispheres. However, in the hemisphere with the positive direction of our movement, we found a tension of $3.8 \sigma$ between two particular redshift bins. All results can be summarized as:

\begin{enumerate}
    \item \textbf{When $z<0.008$ supernovaes are considered, a local decelerated universe in all fits is observed, and tensions $\sim 2\sigma-3\sigma$ between $q_0$ parameters at different redshift and between different distance indicator emerge.} According to \cite{Brout_2022}, the very nearby Hubble Diagram shows a positive bias at $z<0.008$, attributed to measurement errors, unmodeled peculiar velocities and the volumetric bias they induce. Moreover, they suggest to avoid the use of data for $z<0.01$ and also in \cite{Riess_2022} the author neglect data below $z<0.023$ to measurements of the Hubble Flow. Then, is therefore possible that the tensions found in $q_0$ comes from the same origin. However, it is difficult to perform a deep analysis of local large-structure effects if the lowest SNIA redshift are removed, as it is possible that strong effects for the overall cosmology are mislead with this approach. As is presented qualitatively in the \textit{spheres  binned} analysis, the consideration of the lowest redshift SNIA fit a positive value for $q_0$ until $z\sim 0.09$ for the cosmographic luminosity distance. We conclude that it is important to model the very nearby peculiar velocity field to extract cosmological useful information of the most local universe with Pantheon+ as is impossible to do it with the data as it is.
    \item \textbf{Linear regression favours a non-zero slope for $q_0$ in the layer approach with $\Lambda$CDM luminosity distance when the posterior distributions for $q_0$ are assumed to be gaussians.} This result is important because it is observed even removing the lowest redshift supernovae, concluding that the binned values $q_0(z)$ are not uniform across redshift bins. The $\Lambda$CDM model predicts the same value for $q_0$ independent of redshift bin chosen. We consider that a departure from this behaviour can be considered as an evidence against $\Lambda$CDM. This method would become stronger as the quantity of data increase, reducing the errors in the parameter extraction from each redshift binned universe.
    \item \textbf{Not evidence for tensions between UP and DOWN hemispheres.} As previous studies, we did not find evidence for anisotropies between the hemispheres defined by the CMB dipole. In our analysis we used as redshift $z$ the value $z_{HD}$ provided in the catalogue. However, there is an ongoing debate about which is the correct reference frame to be used to find anisotropies in the data. Overall, we do not suggest to use $z_{hel}$ as the frame of reference, because the current cosmology and distance indicators are constructed for observers that are in rest w.r.t. the Hubble flow, requiring a new theoretical approach to analyse the data directly from observed redshift.
    \item \textbf{$3.8 \sigma$ tension for the UP hemisphere at particular redshift bins.} Intriguingly, in one of the hemispheres appears a high tension for the $q_0$ parameter at different redshift bins. If the data is correct, this could be evidence for a local large structure effects and/or peculiar velocities that are not considered in the current cosmological analysis.
\end{enumerate}

\subsection{Interpretations}

We summarized the principal results of our analysis in Pantheon+ survey showing that $q_0$ parameter presents certain anomalies when is compared with what we expect in a full $\Lambda$CDM universe. But, what this deviations could mean for a cosmological perspective?

First, there is the possibility that just locally the universe could be described as a $\Lambda$CDM universe and the global parameters ($\Omega_M \sim 0.3$, $\Omega_\Lambda \sim 0.7$, $\Omega_K \sim 0$) doesn't represent the behaviour of this universe in every place. This point of view is a contradiction: if $\Lambda$CDM model doesn't provide the same parameters in every portion of the universe (in which homogeneity scale is valid), then the universe is not $\Lambda$CDM after all. Different $q_0$ parameters in different places of the universe could indicate that the dynamics of the universe are not the same in every place. We had to remember that our analysis is restricted to the Taylor approximation and a flat $\Lambda$CDM universe, so it is possible that other parameters like curvature $\Omega_k$ are not the same everywhere, biasing the local results on $q_0$. Again, this point of view contradict the cosmological principle, promoting the construction and development of new models. Similar evidence for evolution in cosmological parameter with redshift, namely a decreasing trend for $H_0$ with an increasing one in $\Omega_{m0}$, has been observed in several previous works \cite{Dainotti_2021,Dainotti_22,Colgain_2019,Wong_2019,Millon_2020,Krishnan_2020,_Colg_in_2022,Colgain2022,Camarena_2020}. Those are in agreement with the results shown in Figure \ref{Fit} as $\Omega_{m0}$ is directly related with $q_0$ by Eq. \ref{eq3} and $H_0$ is anti-correlated with $\Omega_{m0}$ in $\Lambda$CDM.

We cannot neglect the idea that the discrepancies observed in the data are attributed to systematics due to the combination of different surveys in the Pantheon sample. According to \cite{Kazantzidis_2020} there are also ``wiggles" in best $\Omega_{m0}$ values for Pantheon data. Presumably, Pantheon+ performs better reducing those systematic anomalies but is possible that still the data has "wiggles". Those systematic effects could potentially explain the anomalies observed in our work, as we compare different surveys at different redshift bins. Also it is necessary to perform the same analysis made in this work over mock data to ensure that the results are not generated by the binning procedure itself. As our analysis consider multiple binning and a lot of different statistical test that requires expensive time computation, we are preparing this analysis for a future work. 

In a different approach, if the universe is not $\Lambda$CDM, then is valid to ask if the $q_0$ value extracted from the cosmological luminosity distance is indeed an acceleration; for example in homogeneous cosmology models as LTB, $q_0$ is product of the radial inhomogeneities of our universe and not a real cinematic quantity. Moreover, some fits of the cosmological Taylor approximations predicts values of $q_0$ out of the range predicted by $\Lambda$CDM cosmology ($q_0 \in [-1,0.5]$), but we need more data to confidently constrain those values.

Also, the possibility that just the data is not well treated has to be considered. There has been wide criticism about the treatment of Supernovae IA data as not considering possible magnitude evolution or the introduction of redshift dependent nuisance parameters in order to account for correlations between properties of SNIA and redshift. Those correlations are attributed just to intrinsic properties of SNIA and not related to cosmology. Finally, we state that any definitive conclusion is far from being strongly stated, as is really important to get much more data in order to constrain better the deviation of $q_0$ from what is expected in $\Lambda$CDM cosmology.

\section{Conclusions}

We proposed two approaches to extract information about the local deceleration parameter $q_0$ from different redshift binned universe, using the well known cosmological Taylor expansion of the luminosity distance and also a flat $\Lambda$CDM universe. The first accounting for a cumulative effect (\textit{spheres}) and the second to study differences in cosmological parameter for different redshift bins (\textit{layers}). We found that when $z<0.008$ supernovae are added to cosmological analysis, tensions $\sim 2\sigma-3\sigma$ emerge between different redshift bins and different luminosity distance indicator. We suggest not to use lowest Supernovae IA in cosmological analysis, but we also state the importance to improve the scientific utility of the data at lower redshift as are necessary to understand the influence of the local large-scale structure in cosmology. Also, using the cosmological Taylor expansion of the luminosity distance, we observe a strong tension $\sim 3.9$ between particular redshift bins, when we analyze the data on the hemisphere that the CMB dipole points. However, significant anisotropies between the 2 hemispheres were not observed. Finally, performing a simple general linear model fit against the redshift binned $\Lambda$CDM parameters, we found that the $q_0(z)$ function prefers a non-zero slope over a model with zero slope expected from a $\Lambda$CDM universe. Those results has to be took carefully as the quantity of data is not that big to state definitive conclusions. We proposed this method as a test for $\Lambda$CDM cosmology, which could be improved as more data is release. 

Finally, we conclude that binned cosmology will be a promising tool to analyze deeply the structure of our universe without assuming the cosmological principle. Our analysis is similar to a recent paper \cite{Perivola:2023} -- that use the Pantheon + sample -- where the authors have shown a $2$ to $3$ $\sigma$ tension between the low and high redshift best fit for the SNIA absolute magnitude $M$. However they did not find variations in $\Omega_{m0}$ and therefore not in $q_0$. As the quantity of data increase, this tool could be performed jointly with anisotropic cosmology in order to get the best insights of the deviations of our universe from $\Lambda$CDM model.

\section*{ACKNOWLEDGMENTS}
EP acknowledge the support through a graduate scholarship ANID-Subdirecci\'on de Capital Humano/Doctorado Nacional/2021-21210824. We wish thank Leandros Perivolaropoulos for his comments on the low redshift volumetric bias on SNIA data.

\appendix
 \section{On the inhomogeneity of redshift bins}

\begin{figure}[H]
    \centering
    \includegraphics[width=8cm]{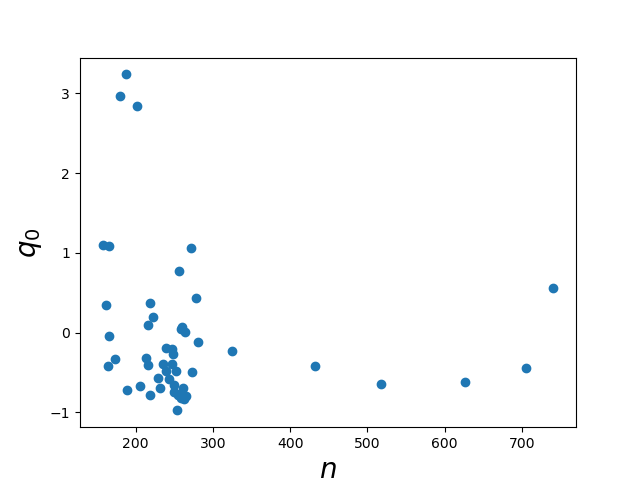}
    \includegraphics[width=8cm]{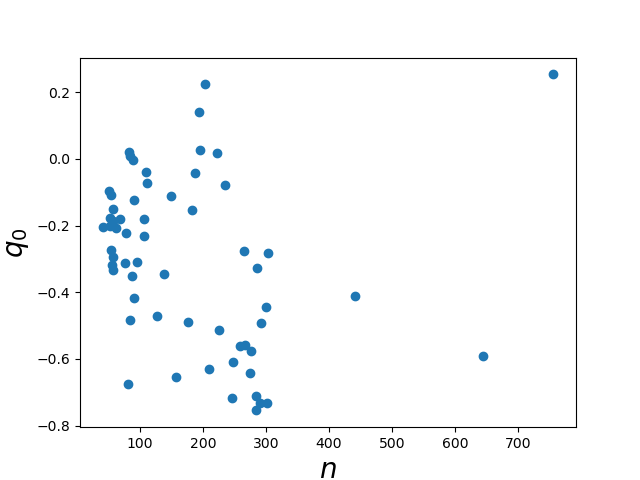}
    \caption{$q_0$ values against the number $n$ of data in each redshift bin for the Cosmographic (left) and the $\Lambda$CDM (right) approach. The correlation coefficients are $R=-0.168$ and $R=-0.201$ respectively.} 
    \label{Correlations}
\end{figure}

\begin{figure}[H]
    \centering
    \includegraphics[width=8cm]{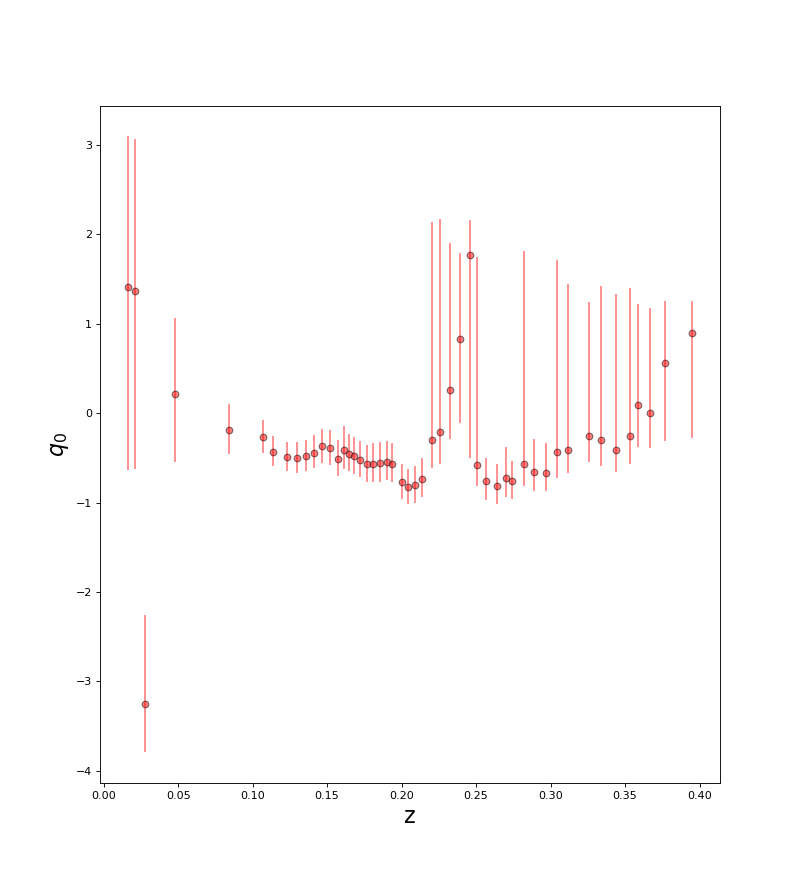}
    \includegraphics[width=8cm]{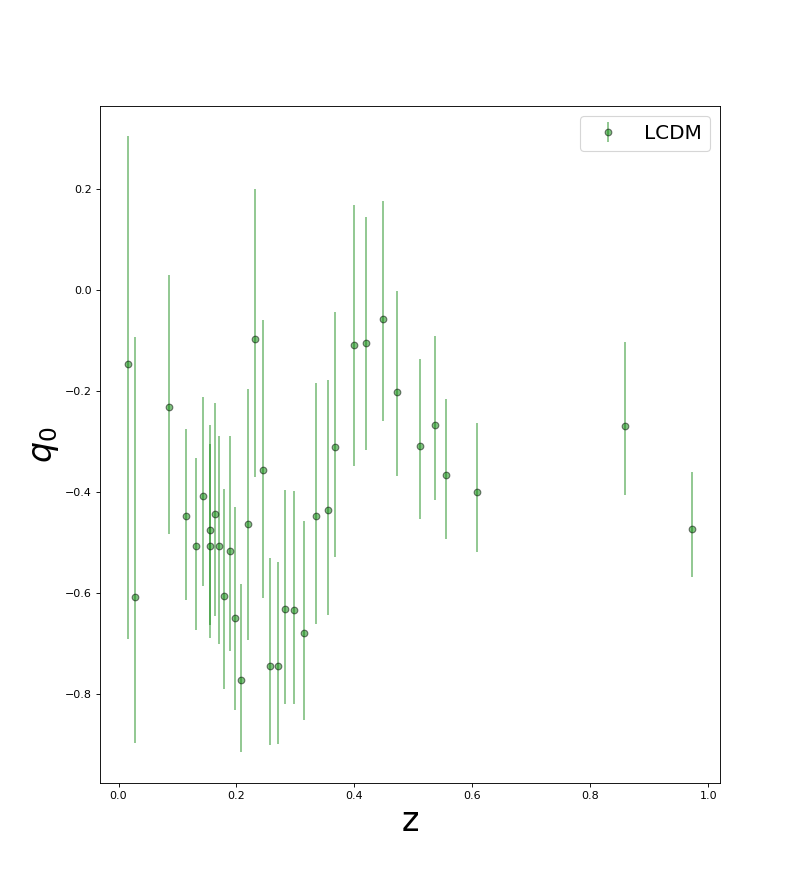}
    \caption{$q_0$ values against $z$ for cosmographic and $\Lambda$CDM luminosity distance in the layer approach using redshift bins with exactly $300$ supernovaes each one. $z$ here is the mean redshift value of the bin.  Note that the bins are not equispaced anymore and the presence of the out-layer value around $z\sim 0.03$ possible attributed to the influence of lower redshift supernovaes even when $z<0.008$ are removed. } 
    \label{Homogenous}
\end{figure}

There is the possibility that the inhomogeneity in the quantity of data in each redshift bin affect our results. To study this concern, we seek for correlations between $q_0$ best values and the quantity of data in each redshift bin for Cosmographic and $\Lambda$CDM approach. Plots of $q_0$ against the quantity of data in each bin $n$ appears in Figure \ref{Correlations} for the 2 approaches. The correlation coefficients are $R=-0.168$ and $R=-0.201$ respectively, which pointed a slightly negative but statistically negligible correlation. We can address the inhomogeneity in bins further studying $q_0$ over redshift bins with $z>0.008$ each one having $300$ supernovaes. Note that despite of an homogeneous distribution of data across redshift bins, each redshift space is not homogeneous anymore. To perform a good comparison, we select the lower limits of the bin to be the same obtained in the layer approach for cosmographic and $\Lambda$CDM luminosity distance, varying therefore the upper limit. Results are plotted in Figure \ref{Homogenous}. 

\begin{figure}[ht]
    \centering
    \includegraphics[width=8.5cm]{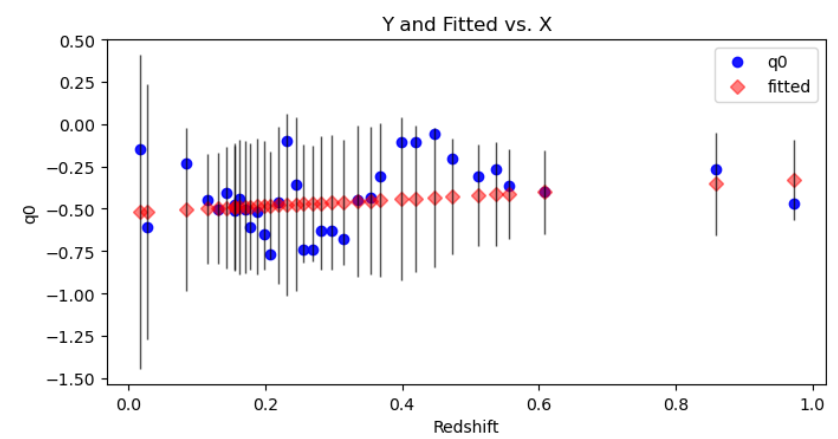}
    \includegraphics[width=8.5cm]{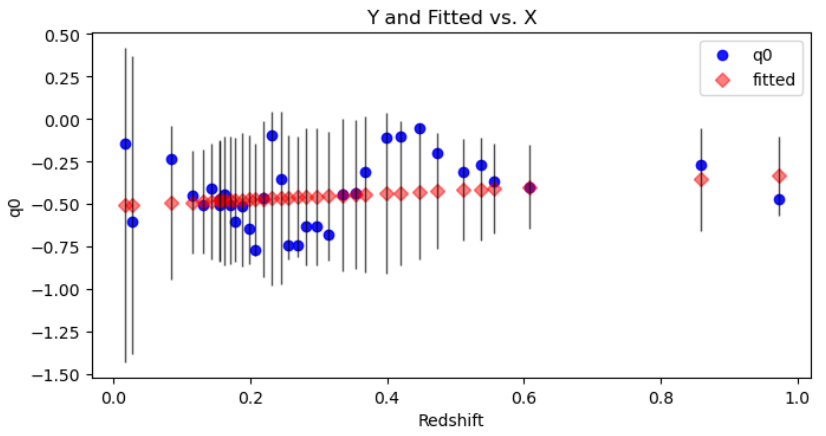}
    \caption{Best general linear model fits for $q_0(z)$ using homogeneous redshift bins, each one with $300$ SNIA. We used $\Lambda$CDM luminosity distance with Gaussian error $\sigma$ estimated as the mean of the asymmetric errors from MCMC (Left) and estimated as the maximum value between them (Right).} 
    \label{Fit2}
\end{figure}

\begin{figure}[h]
    \centering
    \includegraphics[width=10cm]{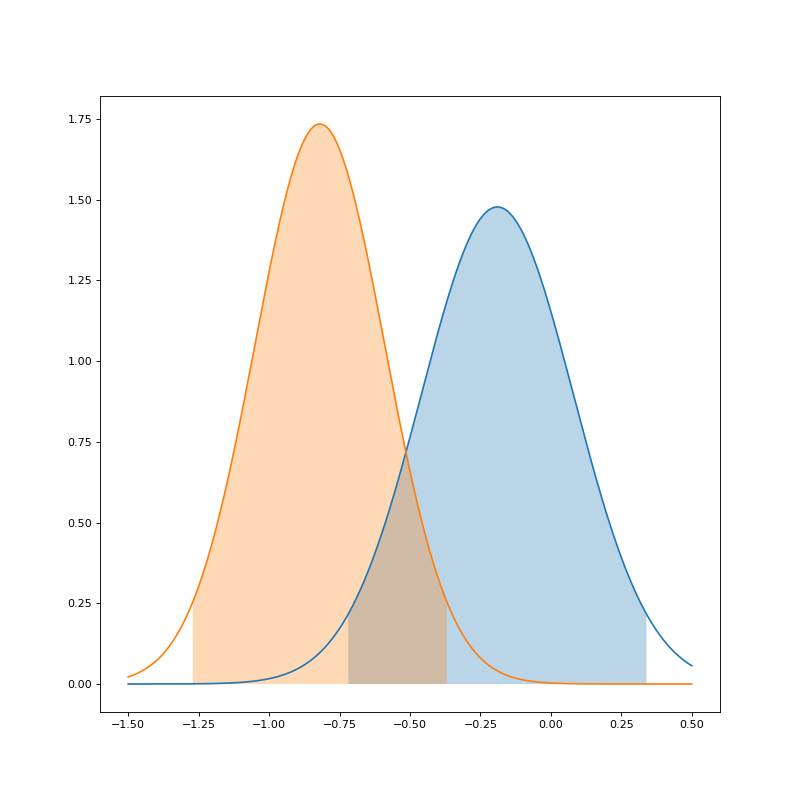}
    \caption{Gaussian distribution of $q_0$ cosmographic parameter for different redshift homogeneous bins, each one of 300 SNIA. Between $[0.0312,0.1364]$ (blue) and $[0.2052,0.3221]$ (orange) appears a tension of $1.78 \sigma$ considering independent realizations.}
    \label{sigmasultima}
\end{figure}

The principal problem with this approach, is that we are not able to study higher redshift data isolated as those bins are extremely wider. For example the last bin with 300 SNIA data is $z \in [0.40,1.54]$, covering more than $\Delta z \sim 1$ space. Indeed, if we try to fit the general linear model over $q_0(z)$ function, we cannot discriminate between zero or non-zero slope. Results are plotted in Figure \ref{Fit2} and best statistics in Table \ref{tablafinal}. Finally, we tried to find tensions at low redshift data with the cosmographic Taylor expansion approach analogous as the results shown in Table \ref{tabla1} and Figure \ref{sigmas1}, but this time using the homogeneous redshift bin. We found a tension of $\sim 1.7\sigma$ between 2 particular redshifts. Results are plotted in Figure \ref{sigmasultima} and details in Table \ref{tablaultima}. Note that SNIA with $z<0.008$ are not influencing this result.

\begin{table}[ht]
\begin{ruledtabular}
\begin{tabular}{lcc}
Statistic & Best linear fit of SNIA & Constrained fit with zero slope \\
\hline
$Q_m$ & $0.1970 \pm 0.110$ ($0.1817 \pm     0.108 $) & 0  \\ 
$Q_0$  & $-0.5220 \pm 0.053  $ ($-0.5109   \pm  0.052$)& $-0.4429     \pm 0.030(-0.4383    \pm 0.030)$ \\ 
Log-Likelihood & $10.061(10.356)$ & $8.4402(8.9112)$  \\
Pearson Chi-Square & $ 27.4(22.9)$  & $30.1(24.8)$  \\
Pseudo R-Square & $0.08897(0.07949)$ & $0.0004160(0.0004160)$ \\
AIC & $-16.12(-16.71)$ & $-14.88(-15.82)$ \\
BIC & $-13.01(1-13.60)$ & $-13.32(-14.27)$
\end{tabular}
\caption{ Best statistics for general linear model fits of $q_0$ using $\Lambda$CDM luminosity distance, estimating the Gaussian error $\sigma$ as the average between asymmetric errors from MCMC.  In parenthesis, same values but this time estimating $\sigma$ as the maximum value between them.}

\label{tablafinal}
\end{ruledtabular}
\end{table}
\begin{table}[ht]
\begin{ruledtabular}
\begin{tabular}{ccc}
 $z$ Bin & Number of SNIA & $q_0 $ \\
\hline
$0.0312-0.1364$  & 300 & $-0.19 \substack{+0.27\\-0.29}$ \\ 
$0.0544-0.20549$ & 300 & $-0.50  \substack{+0.16\\-0.17}$  \\
$0.1066-0.23684 $ & 300  & $-0.52\substack{+0.19\\-0.21}$  \\
$0.2052-0.3221$ & 300 & $-0.815\substack{+0.25\\-0.20}$ 
\end{tabular}
\caption{Values for $q_0$ fit in wide homogeneous redshift bins using the cosmographic Taylor expansion of the luminosity distance.}
\label{tablaultima}
\end{ruledtabular}
\end{table}

\bibliography{mybib}

\end{document}